# The Relativistic Dirac-Morse Green's function

A. D. Alhaidari

*Physics Department, King Fahd University of Petroleum & Minerals, Box 5047,
Dhahran 31261, Saudi Arabia*
E-mail: haidari@mailaps.org

**Abstract:** Using a recently developed approach for solving the three dimensional Dirac equation with spherical symmetry, we obtain the two-point Green's function of the relativistic Dirac-Morse problem. This is accomplished by setting up the relativistic problem in such a way that makes comparison with the nonrelativistic problem highly transparent and results in a mapping of the latter into the former. The relativistic bound states energy spectrum is easily obtained by locating the energy poles of the Green's function components in a simple and straightforward manner.

**Introduction:** Despite all the work that has been done over the years on the Dirac equation, its exact solutions for local interaction has been limited to a very small set of potentials. Since the original work of Dirac in the early part of last century up until 1989 only the relativistic Coulomb problem was solved exactly. In 1989, the relativistic extension of the oscillator problem (Dirac-Oscillator) was finally formulated and solved by Moshinsky and Szczepaniak [1]. Recently, and in a series of articles [2-6], we presented an effective approach for solving the three dimensional Dirac equation for spherically symmetric potential interaction. The first step in the program started with the realization that the nonrelativistic Coulomb, Oscillator, and S-wave Morse problems belong to the same class of shape invariant potentials which carries a representation of so(2,1) Lie algebra. Therefore, the fact that the relativistic version of the first two problems (Dirac-Coulomb and Dirac-Oscillator) were solved exactly makes the solution of the third, in principle, feasible. Indeed, the relativistic Dirac-Morse problem was formulated and solved in Ref. 2. The bound state energy spectrum and spinor wavefunctions were obtained. Taking the nonrelativistic limit reproduces the familiar Schrödinger-Morse problem. Motivated by these findings, the same approach was applied successfully in obtaining solutions for the relativistic extension of yet another class of shape invariant potentials [3]. These included the Dirac-Scarf, Dirac-Rosen-Morse I & II, Dirac-Pöschl-Teller, and Dirac-Eckart potentials. Furthermore, using the same formalism quasi exactly solvable systems at rest mass energies were obtained for a large class of power-law relativistic potentials [4]. Quite recently, Guo Jian-You *et al* succeeded in constructing solutions for the relativistic Dirac-Woods-Saxon and Dirac-Hulthén problems using the same approach [7]. In the fourth and last article of the series in our program of searching for exact solutions to the Dirac equation [5], we found a special graded extension of so(2,1) Lie algebra. Realization of this superalgebra by 2×2 matrices of differential operators acting in the two component spinor space was constructed. The linear span of this graded algebra gives the canonical form of the radial Dirac Hamiltonian. It turned out that the Dirac-Oscillator class, which also includes the Dirac-Coulomb and Dirac-Morse, carries a representation of this supersymmetry.

The central idea in the approach is to separate the variables such that the two coupled first order differential equations resulting from the radial Dirac equation generate Schrödinger-like equations for the two spinor components. This makes the solution of the relativistic problem easily attainable by simple and direct correspondence with well-known exactly solvable nonrelativistic problems. There are two main ingredients in the formulation of the approach that makes it work. The first is a unitary transformation of the Dirac equation which, of course, reduces to the identity in the nonrelativistic limit. The second is the introduction, in a natural way, of an extra potential component which is constrained to depend, in a particular way, on the independent potential function of the problem.

The main objective in all previous applications of the approach was in obtaining the discrete energy spectrum and spinor wavefunctions [2-7]. In this article, however, we demonstrate how one can utilize the same approach in generating the two-point Green's function which is an important object of prime significance in the calculation of physical processes where relativistic effects become relevant. The main contribution here is in obtaining the relativistic Green's function for the Dirac-Morse problem, which is then used to give the relativistic bound states energy spectrum in a simple and direct manner. For completeness and clarity of presentation, we start by giving a brief account of how to construct the Green's function of the nonrelativistic problem by transforming that of another (reference) problem which belongs to the same class. We take the three-dimensional isotropic oscillator as the reference problem and use "point canonical transformation" (PCT) [8,9] to map it into the Green's function for the nonrelativistic S-wave Morse problem. This is possible because, as stated above, the two problems belong to the same class which carries a representation of the dynamical symmetry group SO(2,1).

**Mapping of Green's function under PCT:** The nonrelativistic radial Green's function $\mathcal{G}_\ell(\rho,\rho',\mathcal{E})$ of the three dimensional isotropic oscillator satisfies the following time-independent Schrödinger equation

$$\left[-\frac{d^2}{d\rho^2}+\frac{\ell(\ell+1)}{\rho^2}+\omega^4\rho^2-2\mathcal{E}\right]\mathcal{G}_\ell(\rho,\rho',\mathcal{E})=-2\delta(\rho-\rho') \qquad (1)$$

where $\ell$ is the angular momentum quantum number, $\omega$ is the oscillator frequency, and $\mathcal{E}$ is the nonrelativistic energy. The nonrelativistic Green's function of the S-wave Morse problem, on the other hand, satisfies the following equation [2,9]:

$$\left[-\frac{d^2}{dr^2}+A^2 e^{-2\mu r}-A(2B+\mu)e^{-\mu r}-2E\right]g_\mu(r,r',E)=-2\delta(r-r') \qquad (2)$$

where $\mu$ is the potential range parameter. $A$ and $\mu$ are real and positive. It is to be noted that our definition of the radial Green's function differs by a factor of $(rr')^{-1}$ from other typical definitions. Now, we apply to Eq. (1) the following transformation:

$$\rho=q(r),\ \mathcal{G}_\ell(\rho,\rho',\mathcal{E})=p(r)g_\mu(r,r',E)p^*(r') \qquad (3)$$

If the result is a mapping into equation (2) then this transformation will be referred to as point canonical transformation (PCT). The action of (3), for real functions, on equation (1) maps it into the following equation

$$\left[-\frac{d^2}{dr^2}+\left(\frac{q''}{q'}-2\frac{p'}{p}\right)\frac{d}{dr}+\left(\frac{q''}{q'}\frac{p'}{p}-\frac{p''}{p}\right)+\ell(\ell+1)\left(\frac{q'}{q}\right)^2+\right.$$
$$\left.+(q')^2\left(\omega^4 q^2-2\mathcal{E}\right)\right]g_\mu(r,r',E)=-\frac{2(q')^2}{p(r)p(r')}\delta\big(q(r)-q(r')\big) \qquad (4)$$

where the primes on the transformation functions $p$ and $q$ denote derivatives with respect to $r$. Identifying this with Eq. (2) and using the relation $q'\delta(q(r)-q(r')) = \delta(r-r')$ gives $p(r) = \sqrt{dq/dr}$ and results in the following constraint on the transformation (3) to be a PCT:

$$A^2 e^{-2\mu r} - A(2B+\mu)e^{-\mu r} - 2E = (q')^2\left(\omega^4 q^2 - 2\mathcal{E}\right) + \ell(\ell+1)\left(\frac{q'}{q}\right)^2 + \frac{3}{4}\left(\frac{q''}{q'}\right)^2 - \frac{1}{2}\frac{q'''}{q'} \tag{5}$$

This constraint is solved by taking the PCT function $q(r) = e^{-\mu r/2}$, which will result in the following PCT parameter map:

$$\begin{aligned}\omega^2 &= \tfrac{2A}{\mu} \\ \mathcal{E} &= \tfrac{4A}{\mu}\left(\tfrac{B}{\mu}+\tfrac{1}{2}\right) \\ \ell &= \tfrac{2}{\mu}\sqrt{-2E} - \tfrac{1}{2}\end{aligned} \tag{6}$$

Now, the nonrelativistic radial Green's function for the three dimensional oscillator is well known [10]. It could be written as

$$\mathcal{G}_\ell(\rho,\rho',\mathcal{E}) = \frac{\Gamma\left(\frac{2\ell+3}{4} - \mathcal{E}/2\omega^2\right)}{\omega^2 \Gamma\left(\ell+\frac{3}{2}\right)} \frac{1}{\sqrt{\rho\rho'}} \mathcal{M}_{\mathcal{E}/2\omega^2,\frac{2\ell+1}{4}}(\omega^2 \rho_<^2)\mathcal{W}_{\mathcal{E}/2\omega^2,\frac{2\ell+1}{4}}(\omega^2 \rho_>^2) \tag{7}$$

where $\Gamma$ is the gamma function and $\rho_<(\rho_>)$ is the smaller (larger) of $\rho$ and $\rho'$. $\mathcal{M}_{a,b}$ and $\mathcal{W}_{a,b}$ are the Whittaker functions of the first and second kind, respectively [11]. The two mappings (3) and (6) transform this Green's function into the following one for the nonrelativistic Morse problem [12]:

$$g_\mu(r,r',E) = \frac{\Gamma\left(\frac{1}{\mu}\sqrt{-2E} - B/\mu\right)}{A\,\Gamma\left(1+\frac{2}{\mu}\sqrt{-2E}\right)} e^{\mu(r+r')/2} \mathcal{M}_{\frac{B}{\mu}+\frac{1}{2},\frac{1}{\mu}\sqrt{-2E}}\left(\tfrac{2A}{\mu}e^{-\mu r_>}\right)\mathcal{W}_{\frac{B}{\mu}+\frac{1}{2},\frac{1}{\mu}\sqrt{-2E}}\left(\tfrac{2A}{\mu}e^{-\mu r_<}\right) \tag{8}$$

The switching of arguments of the Whittaker functions is because $\rho_<(\rho_>)$ corresponds to $r_>(r_<)$, respectively. Next, we set up the relativistic problem using the approach mentioned above to obtain the relativistic extension of this Green's function for the Dirac-Morse problem.

**Solving the Dirac equation:** In atomic units ($m = \hbar = 1$) and taking the speed of light $c = \lambdabar^{-1}$, we write the Hamiltonian for a Dirac spinor coupled to a four-component potential $(A_0, \vec{A})$ as follows:

$$H = \begin{pmatrix} 1+\lambdabar A_0 & -i\lambdabar\vec{\sigma}\cdot\vec{\nabla} + i\lambdabar\vec{\sigma}\cdot\vec{A} \\ -i\lambdabar\vec{\sigma}\cdot\vec{\nabla} - i\lambdabar\vec{\sigma}\cdot\vec{A} & -1+\lambdabar A_0 \end{pmatrix} \tag{9}$$

where $\lambdabar$ is the Compton wavelength scale parameter $\hbar/mc$ and $\vec{\sigma}$ are the three 2×2 Pauli matrices. It is to be noted that this type of coupling does not support an interpretation of $(A_0, \vec{A})$ as the electromagnetic potential unless, of course, $\vec{A} = 0$ (e.g., the Coulomb potential). That is, the wave equation with this Hamiltonian is not invariant under the usual electromagnetic gauge transformation. Imposing spherical symmetry and writing $(A_0, \vec{A}) = [\lambdabar V(r), \hat{r}W(r)]$, where $\hat{r}$ is the radial unit vector, gives the following two component radial Dirac equation

$$\begin{pmatrix} 1+\lambdabar^2 V(r)-\varepsilon & \lambdabar\left[\dfrac{\kappa}{r}+W(r)-\dfrac{d}{dr}\right] \\ \lambdabar\left[\dfrac{\kappa}{r}+W(r)+\dfrac{d}{dr}\right] & -1+\lambdabar^2 V(r)-\varepsilon \end{pmatrix} \begin{pmatrix} f^+(r) \\ f^-(r) \end{pmatrix} = 0 \qquad (10)$$

where $\varepsilon$ is the relativistic energy and $\kappa$ is the spin-orbit quantum number defined as $\kappa = \pm(j+\tfrac{1}{2})$ for $\ell = j \pm \tfrac{1}{2}$. $V(r)$ and $W(r)$ are real radial functions referred to as the even and odd components of the relativistic potential, respectively. Eq. (10) results in two coupled first order differential equations for the two radial spinor components. Eliminating one component in favor of the other gives a second order differential equation. This will not be Schrödinger-like (i.e., it contains first order derivatives) unless $V = 0$. To obtain Schrödinger-like equation in the general case we proceed as follows. A global unitary transformation $\mathcal{U}(\eta) = \exp(\tfrac{i}{2}\lambdabar\eta\sigma_2)$ is applied to the Dirac equation (10), where $\eta$ is a real constant parameter and $\sigma_2$ is the 2×2 matrix $\begin{pmatrix} 0 & -i \\ i & 0 \end{pmatrix}$. The Schrödinger-like requirement relates the two potential components by the linear constraint $V(r) = \zeta[W(r)+\kappa/r]$, where $\zeta$ is a real parameter which is related to the transformation parameter $\eta$ by $\sin(\lambdabar\eta) = \pm\lambdabar\zeta$. This results in a Hamiltonian that will be written in terms of only one arbitrary potential function; either the even potential component $V(r)$ or the odd one $W(r)$. The unitary transformation together with the potential constraint map Eq. (10) into the following one, which we choose to write in terms of the even potential component [2,3]

$$\begin{pmatrix} C-\varepsilon+(1\pm 1)\lambdabar^2 V & \lambdabar\left(\mp\zeta+\dfrac{C}{\zeta}V-\dfrac{d}{dr}\right) \\ \lambdabar\left(\mp\zeta+\dfrac{C}{\zeta}V+\dfrac{d}{dr}\right) & -C-\varepsilon+(1\mp 1)\lambdabar^2 V \end{pmatrix} \begin{pmatrix} \phi^+(r) \\ \phi^-(r) \end{pmatrix} = 0 \qquad (11)$$

where $C = \cos(\lambdabar\eta)$ and $\begin{pmatrix} \phi^+ \\ \phi^- \end{pmatrix} = \mathcal{U}\begin{pmatrix} f^+ \\ f^- \end{pmatrix}$. This gives the following equation for one spinor component in terms of the other

$$\phi^{\mp}(r) = \dfrac{\lambdabar}{C\pm\varepsilon}\left[-\zeta\pm\dfrac{C}{\zeta}V(r)+\dfrac{d}{dr}\right]\phi^{\pm}(r) \qquad (12)$$

On the other hand, the resulting Schrödinger-like wave equation for the two spinor components reads

$$\left[-\dfrac{d^2}{dr^2}+\left(\dfrac{C}{\zeta}\right)^2 V^2 \mp \dfrac{C}{\zeta}\dfrac{dV}{dr}+2\varepsilon V-\dfrac{\varepsilon^2-1}{\lambdabar^2}\right]\phi^{\pm}(r) = 0 \qquad (13)$$

In all relativistic problems that have been successfully tackled so far, Eq. (13) is solved by correspondence with well-known exactly solvable nonrelativistic problems [2-7]. This correspondence results in two parameter maps (one for each spinor component) relating the relativistic to the nonrelativistic problem. Using these maps and the known solutions (energy spectrum and wavefunctions) of the nonrelativistic problem one can easily and directly obtain the relativistic energy spectrum and spinor wavefunctions.

**Relativistic Green's function:** Now to the issue at hand – the Green's function. The relativistic 4×4 two-point Green's function $G(\vec{r},\vec{r}',\varepsilon)$ satisfies the inhomogeneous matrix wave equation $(H-\varepsilon)G = -\lambdabar^2\delta(\vec{r}-\vec{r}')$, where the energy $\varepsilon$ does not belong to the spectrum of $H$. For problems with spherical symmetry, the 2×2 radial component $\mathcal{G}_\kappa(r,r',\varepsilon)$ of $G$ satisfies $(H_\kappa-\varepsilon)\mathcal{G}_\kappa = -\lambdabar^2\delta(r-r')$, where $H_\kappa$ is the radial

Hamiltonian operator in Eq. (11). Once again, our definition of the radial component of the Green's function differs by a factor of $(rr')^{-1}$ from other typical definitions. We write $\mathcal{G}_\kappa$ as

$$\mathcal{G}_\kappa(r,r',\varepsilon) = \begin{pmatrix} \mathcal{G}_\kappa^{++} & \mathcal{G}_\kappa^{+-} \\ \mathcal{G}_\kappa^{-+} & \mathcal{G}_\kappa^{--} \end{pmatrix} \tag{14}$$

where $\mathcal{G}_\kappa(r,r',\varepsilon)^\dagger = \mathcal{G}_\kappa(r',r,\varepsilon)$. Let $\Phi = \begin{pmatrix} \phi^+ \\ \phi^- \end{pmatrix}$ and $\overline{\Phi} = \begin{pmatrix} \overline{\phi}^+ \\ \overline{\phi}^- \end{pmatrix}$ be the regular and irregular solutions (at the origin) of Eq. (13), respectively. Using these two solutions, $\mathcal{G}_\kappa$ could be constructed as

$$\mathcal{G}_\kappa(r,r',\varepsilon) = \frac{1}{\Omega_\kappa(\varepsilon)} \left[ \theta(r'-r) \Phi(r,\varepsilon) \overline{\Phi}^\top(r',\varepsilon) + \theta(r-r') \overline{\Phi}(r,\varepsilon) \Phi^\top(r',\varepsilon) \right] \tag{15}$$

where $\theta(r'-r)$ is the Heaviside unit step function and $\Omega_\kappa(\varepsilon)$ is the Wronskian of the regular and irregular solutions:

$$\Omega_\kappa(\varepsilon) = \lambdabar^{-1} \Phi^\top(r,\varepsilon) \begin{pmatrix} 0 & 1 \\ -1 & 0 \end{pmatrix} \overline{\Phi}(r,\varepsilon) = \lambdabar^{-1} \left[ \phi^+(r,\varepsilon) \overline{\phi}^-(r,\varepsilon) - \phi^-(r,\varepsilon) \overline{\phi}^+(r,\varepsilon) \right] \tag{16}$$

which is independent of $r$ as can be verified by differentiating with respect to $r$ and using Eq. (12). Eq. (15) results in the following expressions for the elements of $\mathcal{G}_\kappa$:

$$\mathcal{G}_\kappa^{\pm\pm}(r,r',\varepsilon) = \frac{1}{\Omega_\kappa(\varepsilon)} \phi^\pm(r_<,\varepsilon) \overline{\phi}^\pm(r_>,\varepsilon) \tag{17}$$

$$\mathcal{G}_\kappa^{\pm\mp}(r,r',\varepsilon) = \frac{1}{\Omega_\kappa(\varepsilon)} \left[ \theta(r'-r) \phi^\pm(r,\varepsilon) \overline{\phi}^\mp(r',\varepsilon) + \theta(r-r') \phi^\mp(r',\varepsilon) \overline{\phi}^\pm(r,\varepsilon) \right] \tag{18}$$

The equations satisfied by these elements are obtained from $(H_\kappa - \varepsilon)\mathcal{G}_\kappa = -\lambdabar^2 \delta(r-r')$. They parallel Eqs. (12) and (13) for $\phi^\pm$ and read as follows:

$$\left[ -\frac{d^2}{dr^2} + \left(\frac{C}{\zeta}\right)^2 V^2 \mp \frac{C}{\zeta} \frac{dV}{dr} + 2\varepsilon V - \frac{\varepsilon^2 - 1}{\lambdabar^2} \right] \mathcal{G}_\kappa^{\pm\pm}(r,r',\varepsilon) = -(C \pm \varepsilon) \delta(r-r') \tag{19}$$

$$\mathcal{G}_\kappa^{\mp\pm}(r,r',\varepsilon) = \frac{\lambdabar}{C \pm \varepsilon} \left[ -\zeta \pm \frac{C}{\zeta} V(r) + \frac{d}{dr} \right] \mathcal{G}_\kappa^{\pm\pm}(r,r',\varepsilon) \tag{20}$$

Using the exchange symmetry $r \leftrightarrow r'$ of $\mathcal{G}_\kappa$, Eq. (20) could be rewritten as a linear combination of the $\mathcal{G}_\kappa^{\pm\pm}$ terms with two real coefficients adding up to unity. That is, we write

$$\mathcal{G}_\kappa^{-+}(r,r',\varepsilon) = \mathcal{G}_\kappa^{+-}(r',r,\varepsilon) = \xi \frac{\lambdabar}{C+\varepsilon} \left[ -\zeta + \frac{C}{\zeta} V(r) + \frac{d}{dr} \right] \mathcal{G}_\kappa^{++}(r,r',\varepsilon)$$
$$+ (1-\xi) \frac{\lambdabar}{C-\varepsilon} \left[ -\zeta - \frac{C}{\zeta} V(r') + \frac{d}{dr'} \right] \mathcal{G}_\kappa^{--}(r,r',\varepsilon) \tag{21}$$

where $\xi$ is an arbitrary real dimensionless parameter. This development will now be applied to our problem.

**Dirac-Morse Green's function:** In this setting, the Dirac-Morse problem is the system described by Eq. (11) with $V(r) = -\mathcal{B}e^{-\mu r}$ and $C = \mathcal{A}/\tau$, where $\tau \mathcal{A} > 0$ [2,5,6]. The parameters are related by $\tau^2 = \mathcal{A}^2 + \lambdabar^2 \mathcal{B}^2$ with $\zeta = \mathcal{B}/\tau$. Consequently, Eq. (19) for the diagonal elements of the radial Green's function reads:

$$\left[-\frac{d^2}{dr^2}+\mathcal{A}^2 e^{-2\mu r}-\mathcal{A}(2\varepsilon\mathcal{B}/\mathcal{A}\pm\mu)e^{-\mu r}-\frac{\varepsilon^2-1}{\hbar^2}\right]\mathcal{G}_\mu^{\pm\pm}(r,r',\varepsilon)=-(\mathcal{A}/\tau\pm\varepsilon)\delta(r-r') \quad (22)$$

Comparing this with Eq. (2) gives the following two maps between the relativistic and non-relativistic problems. The map concerning $\mathcal{G}_\mu^{++}$ is

$$\begin{aligned}
&g_\mu = 2\mathcal{G}_\mu^{++}/(\mathcal{A}/\tau+\varepsilon)\\
&A = \mathcal{A} \quad \text{or} \quad A = -\mathcal{A}\\
&B = \varepsilon\mathcal{B}/\mathcal{A} \quad \text{or} \quad B = -\varepsilon\mathcal{B}/\mathcal{A}-\mu\\
&E = (\varepsilon^2-1)/2\hbar^2
\end{aligned} \quad (23)$$

The choice $B=\varepsilon\mathcal{B}/\mathcal{A}$ or $B=-\varepsilon\mathcal{B}/\mathcal{A}-\mu$ depends on whether $\mathcal{A}>0$ or $\mathcal{A}<0$, respectively. On the other hand, the map for $\mathcal{G}_\mu^{--}$ is as follows:

$$\begin{aligned}
&g_\mu = 2\mathcal{G}_\mu^{--}/(\mathcal{A}/\tau-\varepsilon)\\
&A = \mathcal{A} \quad \text{or} \quad A = -\mathcal{A}\\
&B = \varepsilon\mathcal{B}/\mathcal{A}-\mu \quad \text{or} \quad B = -\varepsilon\mathcal{B}/\mathcal{A}\\
&E = (\varepsilon^2-1)/2\hbar^2
\end{aligned} \quad (24)$$

Similarly, the choice $B=\varepsilon\mathcal{B}/\mathcal{A}-\mu$ or $B=-\varepsilon\mathcal{B}/\mathcal{A}$ depends on whether $\mathcal{A}$ is positive or negative, respectively. The two mappings (23) and (24) transform the nonrelativistic Green's function (8) into the following solutions of Eq. (22):

$$\mathcal{G}_\mu^{++} = \frac{\frac{\mathcal{A}}{\tau}+\varepsilon}{2|\mathcal{A}|}\frac{e^{\mu(r+r')/2}}{\Gamma(1+2\beta)}\begin{cases}\Gamma(\beta-\alpha)\mathcal{M}_{\alpha+\frac{1}{2},\beta}(\frac{2\mathcal{A}}{\mu}e^{-\mu r_>})\mathcal{W}_{\alpha+\frac{1}{2},\beta}(\frac{2\mathcal{A}}{\mu}e^{-\mu r_<}) &, \mathcal{A}>0\\ \Gamma(1+\beta+\alpha)\mathcal{M}_{-\alpha-\frac{1}{2},\beta}(\frac{-2\mathcal{A}}{\mu}e^{-\mu r_>})\mathcal{W}_{-\alpha-\frac{1}{2},\beta}(\frac{-2\mathcal{A}}{\mu}e^{-\mu r_<}) &, \mathcal{A}<0\end{cases} \quad (25)$$

$$\mathcal{G}_\mu^{--} = \frac{\frac{\mathcal{A}}{\tau}-\varepsilon}{2|\mathcal{A}|}\frac{e^{\mu(r+r')/2}}{\Gamma(1+2\beta)}\begin{cases}\Gamma(1+\beta-\alpha)\mathcal{M}_{\alpha-\frac{1}{2},\beta}(\frac{2\mathcal{A}}{\mu}e^{-\mu r_>})\mathcal{W}_{\alpha-\frac{1}{2},\beta}(\frac{2\mathcal{A}}{\mu}e^{-\mu r_<}) &, \mathcal{A}>0\\ \Gamma(\beta+\alpha)\mathcal{M}_{-\alpha+\frac{1}{2},\beta}(\frac{-2\mathcal{A}}{\mu}e^{-\mu r_>})\mathcal{W}_{-\alpha+\frac{1}{2},\beta}(\frac{-2\mathcal{A}}{\mu}e^{-\mu r_<}) &, \mathcal{A}<0\end{cases} \quad (26)$$

where $\alpha=\varepsilon\mathcal{B}/\mu\mathcal{A}$ and $\beta=\frac{1}{\mu\hbar}\sqrt{1-\varepsilon^2}$. The off-diagonal elements of $\mathcal{G}_\mu$ are obtained by substituting these in Eq. (21), which could be rewritten in terms of the variable $x=\frac{2\mathcal{A}}{\mu}e^{-\mu r}$ as

$$\mathcal{G}_\mu^{-+}(x,x',\varepsilon) = \mathcal{G}_\mu^{+-}(x',x,\varepsilon) = -\xi\frac{\hbar\mu}{\frac{\mathcal{A}}{\tau}+\varepsilon}\frac{1}{\sqrt{xx'}}\left(x\frac{d}{dx}+\frac{x}{2}+\frac{\mathcal{B}}{\mu\tau}-\frac{1}{2}\right)\sqrt{xx'}\mathcal{G}_\mu^{++}$$
$$-(1-\xi)\frac{\hbar\mu}{\frac{\mathcal{A}}{\tau}-\varepsilon}\frac{1}{\sqrt{xx'}}\left(x'\frac{d}{dx'}-\frac{x'}{2}+\frac{\mathcal{B}}{\mu\tau}-\frac{1}{2}\right)\sqrt{xx'}\mathcal{G}_\mu^{--} \quad (27)$$

Using the differential formulas of the Whittaker functions [11] we obtain the following expressions for the off-diagonal elements of the Dirac-Morse Green's function

$$\mathcal{G}_\mu^{-+}(r,r',\varepsilon) = \mathcal{G}_\mu^{+-}(r',r,\varepsilon) = \frac{\hbar\mathcal{B}}{\mathcal{A}}\left[(\xi-1)\mathcal{G}_\mu^{--}-\xi\mathcal{G}_\mu^{++}\right]+e^{\mu(r+r')/2}\times$$
$$\left(\xi-\frac{1}{2}\right)\frac{\hbar\mu}{\mathcal{A}}\frac{\Gamma(1+\beta-\alpha)}{\Gamma(1+2\beta)}\left[-\theta(r-r')\mathcal{M}_{\alpha-\frac{1}{2},\beta}(\tfrac{2\mathcal{A}}{\mu}e^{-\mu r})\mathcal{W}_{\alpha+\frac{1}{2},\beta}(\tfrac{2\mathcal{A}}{\mu}e^{-\mu r'})\right.$$
$$\left.+(\beta+\alpha)\theta(r'-r)\mathcal{M}_{\alpha+\frac{1}{2},\beta}(\tfrac{2\mathcal{A}}{\mu}e^{-\mu r'})\mathcal{W}_{\alpha-\frac{1}{2},\beta}(\tfrac{2\mathcal{A}}{\mu}e^{-\mu r})\right] \quad , \mathcal{A}>0 \quad (28)$$

$$\mathcal{G}_\mu^{-+}(r,r',\varepsilon) = \mathcal{G}_\mu^{+-}(r',r,\varepsilon) = \frac{\hbar\mathcal{B}}{\mathcal{A}}\left[(\xi-1)\mathcal{G}_\mu^{--} - \xi\mathcal{G}_\mu^{++}\right] + e^{\mu(r+r')/2} \times$$

$$\left(\xi-\tfrac{1}{2}\right)\frac{\hbar\mu}{\mathcal{A}}\frac{\Gamma(1+\beta+\alpha)}{\Gamma(1+2\beta)}\left[(\beta-\alpha)\theta(r-r')\mathcal{M}_{-\alpha+\frac{1}{2},\beta}(\tfrac{-2\mathcal{A}}{\mu}e^{-\mu r})\mathcal{W}_{-\alpha-\frac{1}{2},\beta}(\tfrac{-2\mathcal{A}}{\mu}e^{-\mu r'})\right., \quad \mathcal{A}<0 \quad (29)$$

$$\left. -\theta(r'-r)\mathcal{M}_{-\alpha-\frac{1}{2},\beta}(\tfrac{-2\mathcal{A}}{\mu}e^{-\mu r'})\mathcal{W}_{-\alpha+\frac{1}{2},\beta}(\tfrac{-2\mathcal{A}}{\mu}e^{-\mu r})\right]$$

where $\xi \neq \tfrac{1}{2}$. One can easily verify that the relativistic bound states energy spectrum of the Dirac-Morse problem [2,5,6] is located at the energy poles of these components of the Green's function. This is simply and directly obtained by taking the argument of the gamma function in the numerator to be equal to $-n$, where $n = 0,1,2\ldots$ That is by taking $\beta - \alpha = -n$ for $\mathcal{A} > 0$ and $\beta + \alpha = -n$ for $\mathcal{A} < 0$ giving the following spectrum:

$$\varepsilon_n = \frac{\mathcal{A}}{\tau^2}\left[\mu\hbar^2\mathcal{B}n \pm \sqrt{\tau^2 - (\mu\hbar\mathcal{A}n)^2}\right] \quad ; n = 0,1,2,\ldots,n_{max} \quad (30)$$

where $n_{max} \leq \frac{1}{\mu\hbar}\sqrt{1+(\hbar\mathcal{B}/\mathcal{A})^2}$.

It might be worthwhile looking at the nonrelativistic limit ($\hbar \to 0$) of the Green's function. One can easily show that in this limit:

$$\varepsilon \approx 1 + \hbar^2 E, \quad \alpha \approx \frac{\mathcal{B}}{\mu\mathcal{A}}\left(1+\hbar^2 E\right), \quad \beta \approx \frac{1}{\mu}\sqrt{-2E}, \quad \tau \approx \mathcal{A} + \hbar^2\mathcal{B}^2/2\mathcal{A} \quad (31)$$

Substituting these in formulas (25) and (26) for the diagonal elements of the Green's function shows that their behavior in the limit is $\mathcal{G}_\mu^{++} \approx g_\mu$ and $\mathcal{G}_\mu^{--} \approx \hbar^2 g_\mu$. On the other hand, the off-diagonal elements $\mathcal{G}_\mu^{\pm\mp}$ go to the limit like $\hbar$, except for the first term in formulas (28) and (29) which is proportional to $\hbar\mathcal{G}_\mu^{--}$. This term goes to the limit like $\hbar^3$. Therefore, the relativistic behavior of the 2×2 radial Green's function could be written symbolically as

$$\mathcal{G}_\mu \sim \begin{pmatrix} 1 & \hbar + \hbar^3 \\ \hbar + \hbar^3 & \hbar^2 \end{pmatrix} \quad (32)$$

We give one final comment concerning the constant parameter $\xi$ that appears in the expressions for $\mathcal{G}_\mu^{\pm\mp}$. Aside from the restriction that $\xi \neq \tfrac{1}{2}$, we found no obvious criterion for the selection of a specific value to be assigned to $\xi$. This issue might be settled based on the outcome of relativistic calculations. It is, however, unfortunate that this author does not have the calculation tools nor the necessary computational skills needed to carry out such a project, which may prove to be very fruitful.

**Acknowledgements:** The author is grateful to Dr. F. A. Al-Sulaiman (KFUPM) for the valuable support in literature survey.